\documentclass[usenatbib]{emulateapj}

\usepackage{graphicx}

\catcode`\@=11
\def\gsim{\ifmmode{\mathrel{\mathpalette\@versim>}}
    \else{$\mathrel{\mathpalette\@versim>}$}\fi}
\def\lsim{\ifmmode{\mathrel{\mathpalette\@versim<}}
    \else{$\mathrel{\mathpalette\@versim<}$}\fi}
\def\@versim#1#2{\lower 2.9truept \vbox{\baselineskip 0pt \lineskip
    0.5truept \ialign{$\m@th#1\hfil##\hfil$\crcr#2\crcr\sim\crcr}}}
\catcode`\@=12

\def\dg{^\circ}

\def\apj{ApJ}
\def\apjl{ApJ}
\def\apjs{ApJS}
\def\aap{A\&A}
\def\mnras{MNRAS}

\begin{document}     

\title{The inner Galactic bulge: evidence for a nuclear bar?}  
\shorttitle{Structure of the inner Galactic bulge-bar}

\author{Ortwin Gerhard\altaffilmark{1} and Inma Martinez-Valpuesta\altaffilmark{1} }
\affil{Max-Planck-Institut f\"ur Extraterrestrische Physik, Giessenbachstrasse, 85748 Garching, Germany}
\shortauthors{Ortwin~Gerhard \& Inma~Martinez-Valpuesta}



\begin{abstract}
  Recent data from the VVV survey have strengthened evidence for a
  structural change in the Galactic bulge inwards of $\vert
  l\vert\le4\dg$.  Here we show with an N-body barred galaxy
  simulation that a boxy bulge formed through the bar and buckling
  instabilities effortlessly matches measured bulge longitude profiles
  for red clump stars.  The same simulation snapshot was earlier used
  to clarify the apparent boxy bulge - long bar dichotomy, for the
  same orientation and scaling.  The change in the slope of the model
  longitude profiles in the inner few degrees is caused by a
  transition from highly elongated to more nearly axisymmetric
  isodensity contours in the inner boxy bulge. This transition is
  confined to a few degrees from the Galactic plane, thus the change
  of slope is predicted to disappear at higher Galactic latitudes. We
  also show that the nuclear star count map derived from this
  simulation snapshot displays a longitudinal asymmetry similar to
  that observed in the 2MASS data, but is less flattened to the
  Galactic plane than the 2MASS map. These results support the
  interpretation that the Galactic bulge originated from disk
  evolution, and question the evidence advanced from star count data
  for the existence of a secondary nuclear bar in the Milky Way.
\end{abstract}

\section{Introduction}   
\label{sec:intro}   

The barred nature of the Milky Way is well-established from NIR
photometry \citep{Blitz+Spergel91, Dwek+95, Binney+97,
  Bissantz+Gerhard02} and star counts \citep{Stanek+94,
  LopezCorredoira+05,Benjamin+05,Cabrera-Lavers+07b}, from comparing
HI and CO lv-diagrams with hydrodynamic models
\citep{Englmaier+Gerhard99, Fux99, Bissantz+03}, and from dynamical
modelling of the kinematics of bulge stars
\citep{Zhao96,Fux97,Bissantz+04,Shen+10}.

One open issue is the structure of the inner Galactic bulge, for
$R\lsim 0.5$ kpc. Is this simply the scaled-down, higher-density parts
of the surrounding boxy bulge? Does it become axisymmetric towards the
center? Does it contain a ``classical'' bulge component? Or does it
even contain a secondary nuclear bar as do about one third of barred
galaxies like ours \citep{Erwin11}? 

This issue is important; for example, a nuclear bar misaligned with
the main Galactic boxy-bulge bar would substantially modify the
morphology of the gas flow in the central Galaxy, as illustrated by
the simulations of \citet[][hereafter
RFC08]{Rodriguez-Fernandez+Combes08} and \citet{Namekata+09}.

\citet{Nishiyama+05} analyzed red clump (RC) star counts in the inner
bulge. Because RC stars are approximate standard candles, their
magnitude distribution is a proxy for the stellar density along the
line-of-sight \citep{Stanek+94}.  Determining the maximum of the RC
magnitude distribution at various longitudes in a strip at $b=-1\dg$,
\citet{Nishiyama+05} found a clear change of slope in the RC longitude
profile at $\vert l\vert=4\dg$, separating a steeper slope in the
outer bulge from a shallow slope in the nuclear regions.

Recently, \citet{Gonzalez+11b} analyzed new RC star counts from the VVV
survey and found excellent agreement with \citet{Nishiyama+05}. They also
determined the RC maximum versus longitude for $b=+1\dg$ and found
good agreement with the $b=-1\dg$ results, as expected for a
triaxially symmetric structure. This also shows that reddening
corrections, which differ greatly between these latitudes, are
unlikely to cause the observed change of slope in the longitude
profiles.

The RC star counts therefore indicate a significant and fairly sudden
change in the structure of the bulge at $\vert l\vert\simeq 4\dg$.
Here we analyze the distribution of RC maxima for a suitably oriented
boxy bulge - barred galaxy model. Surprisingly, we find that the RC
longitude profiles predicted by this model are very similar to the
observed profiles.  We show that this can be traced back to the
model's density structure becoming more axisymmetric towards the
center, and finally we compare the asymmetry of the predicted star
count map in the inner bulge region with available analysis of 2MASS
data.


\section{A Milky Way-like N-body model }
\label{sec:theory}

\begin{figure}
\hspace{20pt}
\includegraphics[angle=-90.0,width=0.75\hsize]{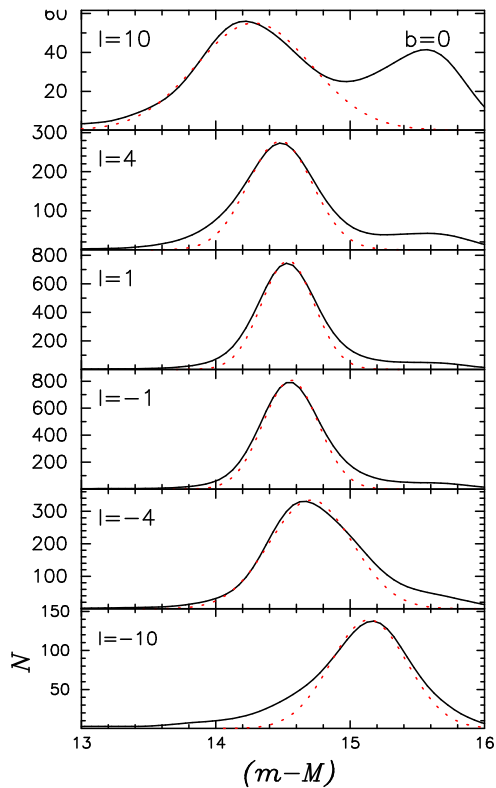}
\caption[]{Distributions of simulated red clump distance moduli in
  different bulge fields (black lines), for an N-body model of the
  Galactic boxy bulge and bar. The model is viewed from a position
  $R_0=8 {\rm kpc}$ from the Galactic center and $\alpha_{\rm
    bar}=25\dg$ away from the bar major axis. The particle positions
  are convolved with a Gaussian luminosity function for the red clump
  stars with width $\sigma_{\rm RC}=0.17$ mag. Gaussian fits to the
  distributions are shown by dotted red lines.}
\label{fig:RCsmoohists}
\end{figure}

The model analyzed in this paper stems from the barred galaxy
simulation of \citet{Martinez-Valpuesta+06} and is the same model
which we used earlier in order to argue that the long bar and the boxy
bulge in the MW are the three-dimensional and planar components of the
MW's main bar \citep[][hereafter MVG11]{Martinez-Valpuesta+Gerhard11}.
It evolved from an initially exponential disk with $Q=1.5$ embedded in
a dark matter halo, and developed a prominent boxy bulge through a
buckling instability after $\sim1.5$ Gyr. To resolve the nuclear bulge
better the particle distribution is symmetrized with respect to the
midplane.  We consider the simulated galaxy at time $\sim1.9$ Gyr,
some time after the instability when the bar has resumed its evolution
and has regrown through further angular momentum transfer to the
halo. The snapshot chosen for our analysis in this paper is that shown
in Fig.~1 of MVG11; however, we have checked that the results given
for the nuclear regions below do not significantly change when the
snapshot in Fig.~2 of MVG11 is used instead.

\subsection{Magnitude-dependent simulated star counts in the inner bulge}
\label{ssec:modelcounts}

\begin{figure}
\hspace{20pt}
\includegraphics[angle=-90.0,width=0.75\hsize]{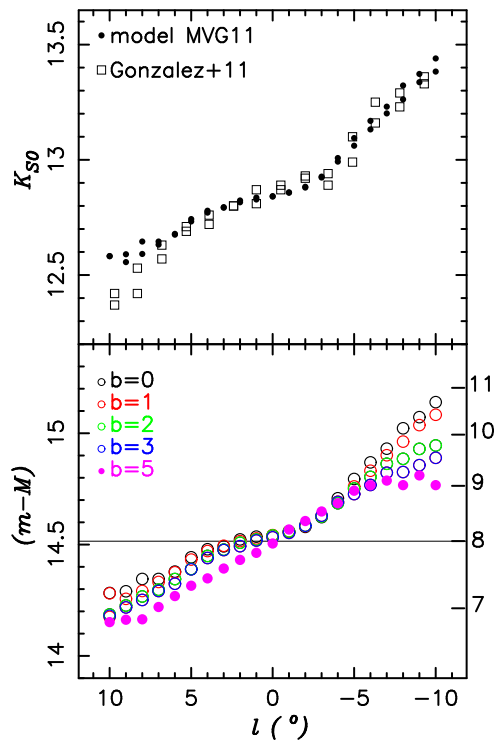}
\caption[]{Maxima of observed and model magnitude distributions for
  red clump (RC) giant stars in bulge fields as a function of
  longitude. Top: Simulated RC maxima for strips with latitudes
  $b=[1\pm0.5\dg]$ and $[0\pm0.5\dg]$ (black dots), compared with data
  from the VVV survey at $b=\pm1\dg$ \citep[][open
  squares]{Gonzalez+11b}.  We used $M_{\rm K}=-1.70$ to shift the model
  distance moduli to the magnitude scale of the data; see
  text. Bottom: RC maxima in strips with different latitudes,
  colour-coded as indicated on the plot. Individual fields have sizes
  $\Delta l=\Delta b=1\dg$ and are separated in longitude by $1\dg$.
  The change of slope in the inner few degrees seen at low latitudes
  is absent at $b=5\dg$.  The horizontal line illustrates the assumed
  distance to the GC.}
\label{fig:RCmaxwithl}
\end{figure}

\begin{figure*}
\hspace{10pt}
\includegraphics[angle=-90.0,width=0.40\hsize]{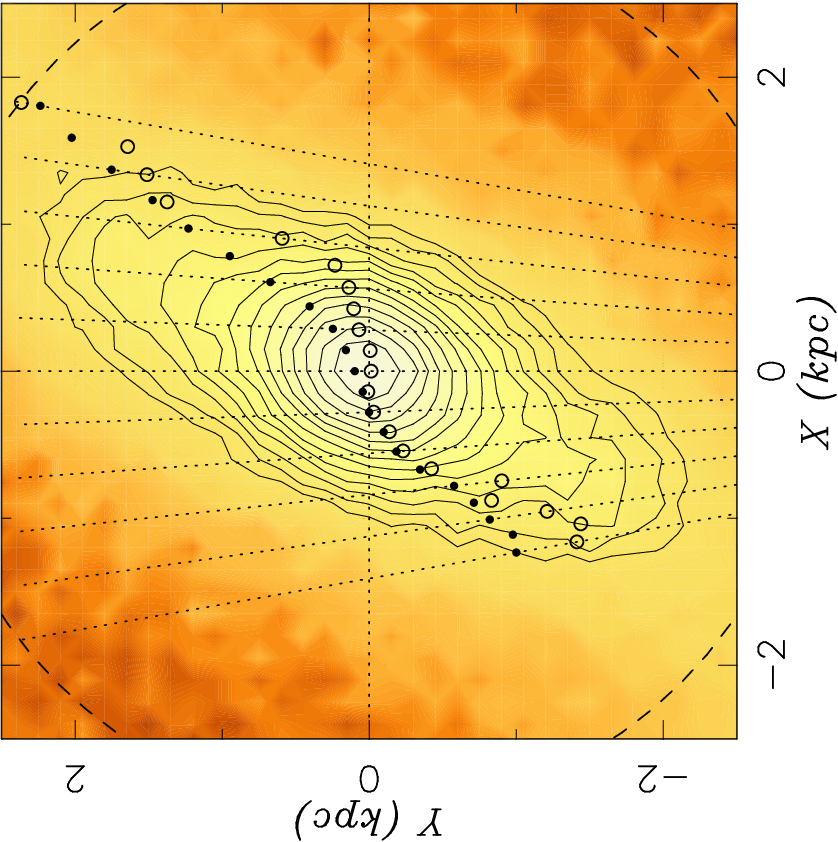}
\hfill
\includegraphics[angle=-90.0,width=0.40\hsize]{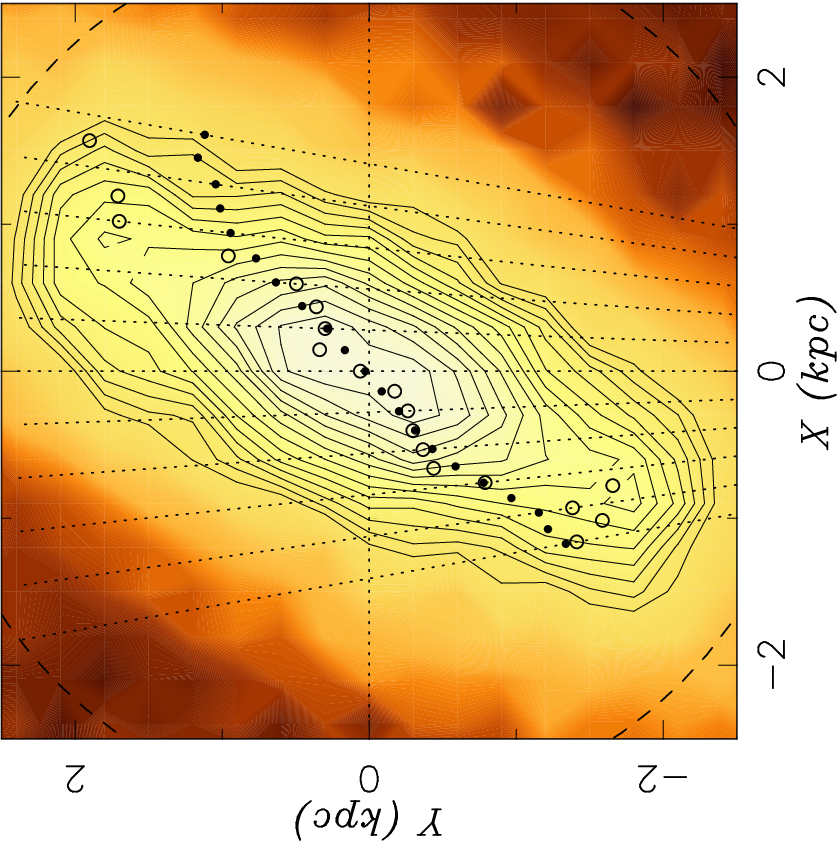}
\hspace{30pt}
\caption[]{\textit{Left panel}: Face-on surface density of the
  particles with $\vert z \vert < 300$ pc, with overplotted maxima of
  the line-of-sight density distributions (open circles) and maxima of
  the simulated line-of-sight RC magnitude distributions (full
  circles) for particles in the latitude range $\vert b\vert \le2\dg$.
  \textit{Right panel}: Same for the particles with $450 {\rm pc} <
  \vert z \vert < 750$ pc, and the latitude range $3\dg<b<5\dg$.  The
  long axis of the bar is at an angle $\alpha_{\rm bar}=25\dg$
  relative to the line-of sight to the observer at the assumed
  galactocentric radius $R_0=8$ kpc.  Dotted lines show directions
  $l=0, \pm2,\pm 4, \pm 6, \pm8, \pm10\dg$, as seen from the
  observer's position.  }
\label{fig:projections}
\end{figure*}

The stellar population of the Galactic bulge is predominantly old
($\sim 10$ Gyr) and has a metallicity distribution with FWHM $\sim 1$
dex around [Fe/H]$\simeq$-0.1 dex \citep{Brown+10}.  In this range the
number of RC stars per unit mass varies within $\sim10\%$
\citep{Salaris+Girardi02}, so we can assume that the spatial
distribution of RC stars is the same as that of model particles
(mass).
We project the bulge of this model onto the sky in longitude-latitude
coordinates $(l,b)$ as seen by an observer at $R_0=8$ kpc distance
from the Galactic center (GC hereafter), whose line-of-sight (LOS) is 
at an angle $\alpha_{\rm bar}=25\dg$ relative to the long axis of the bar
\citep[][MVG11]{Gerhard02}. Dividing the projected particle
distribution into fields of $\Delta l= \Delta b=1\dg$, we sort the
particles in each field by distance modulus. To account for the finite
width of the luminosity function for red clump giant stars (RC
hereafter), we convolve each particle with a Gaussian of width $\sigma=0.17$
mag \citep{Alves00}. The absolute M$_K$(RC) varies by $\sim0.26$ mag
over the observed FWHM metallicity range \citep{Salaris+Girardi02},
corresponding to a $\sigma$ of 0.11 mag. Added in quadrature, this
gives a total $\sigma'=0.21$ mag.  Using $\sigma'$ instead of
$\sigma=0.17$ mag has a negligible effect on the figures shown
below.

Fig.~\ref{fig:RCsmoohists} shows the resulting LOS magnitude
distributions for fields with $b=0$ and several $l$.  The main peak
corresponds to the inclined bar and boxy bulge; it shifts from
m-M$\simeq 14.28$ at $l=10\dg$ on the bar's near side to
m-M$\simeq15.14$ at $l=-10\dg$ on the far side.  The background disk
is seen clearly at m-M$\simeq 15.55$ in the top panel for $l=10\dg$;
the slope towards m-M=$16$ is consistent with the disk
scale-length. This disk feature is similar in absolute numbers but
weaker in relative numbers for smaller $\vert l\vert$ and seems to
blend with the far end of the bar at $l=-10\dg$.  A similar
phenomenology appears to be seen in the VVV data
\citep[][Fig.~2]{Gonzalez+11b}, which in addition show a 'background'
of K giants not included in the Gaussian RC luminosity function.

We determine the maxima of the simulated RC distribution in all fields
by fitting a Gaussian to the brightest peak of the distribution.
Fig.~\ref{fig:RCmaxwithl} shows the resulting RC longitude profiles
for several latitude strips, and compares those at $b=0\dg$ and
$b=1\dg$ to the results from \citet{Gonzalez+11b} at $b=\pm1\dg$. This
last step requires choosing an absolute K magnitude for the RC. To
shift the model to the magnitude scale of the data, we need to use
M$_K$(RC)= -1.70.  The most recent M$_K$(RC)=-1.61 \citep{Alves00,
    Laney+11} for the local RC, corrected for population effects
  \citep{Salaris+Girardi02}, would predict M$_K$(RC)$\simeq-1.52$.  The
  -1.70 value would thus argue for a shorter distance $R_0$ to the
GC, while the best current value is $R_0=8.3\pm0.23$ kpc
\citep{Brunthaler+11}. We do not pursue this issue further here.

Both the observed $b=\pm1\dg$ longitude profiles and the low-latitude
model longitude profiles show the same qualitative result: the steep
slope of the RC maxima seen at large positive and negative $\vert
l\vert > 4\dg$ flattens for small $\vert l\vert$\footnote{Due to the
  symmetrization the model star counts at $b=1\dg$ and $b=-1\dg$ are
  equivalent.}. The model profile is slightly flatter than the
  data inside $l=\pm2\dg$ and at $l>4\dg$, but steeper around
  $l=-4\dg$. The origin of the difference at $l>6\dg$ between model
  and data is unclear; the maxima in the observed distributions are
  very broad and the contribution from the disk substantial, but also
  differential population effects might contribute.  By contrast
  with the lower latitudes, the $b=5\dg$ model longitude profile
shows a markedly weaker flattening in the inner parts $\vert l\vert <
4\dg$.

This model was not constructed to match Milky Way observations -- it
is a generic boxy bulge and bar model from a simulation; thus we do
not expect a perfect match with any Milky Way data. The agreement
between the low-latitude observed and model profiles is therefore even
more striking. Previously, the flattening of the observed longitude
profiles has been interpreted as due to a distinct structure in the
inner Galactic bulge \citep{Nishiyama+05, Gonzalez+11b}, perhaps a
nuclear bar \citep[see also][RFC08]{Alard01}. The simulated model
bulge does not contain such a structure.

\subsection{Origin of the change of slope in the simulated star counts}
\label{ssec:issues}

To understand the cause of the flattening in the model longitude
profiles, let us first consider a simple thought experiment. We know
that the RC maxima for a thin bar follow approximately the major axis
of the bar \citep[e.g.][]{Cabrera-Lavers+07b}.  For a thicker bar,
e.g., with planar axis ratio $b/a\sim 0.5$, a volume effect arises due
to the geometry of the cones with fixed $\Delta l, \Delta b$. This
causes the maxima of the RC magnitude distributions to lie behind the
maxima of the line-of-sight (LOS) density distributions
\citep[][MVG11]{Cabrera-Lavers+07b}. However, as long as the density
distribution is approximately scale-free, we would expect the RC
longitude profile to have a similar slope as for the thin bar,
following the orientation of the major axis.

Now insert an axisymmetric component with high stellar density in the
center of the bar. If its density is sufficiently high, it will
dominate the surrounding bar, and since it is axisymmetric, the slope
of its longitude profile will be nearly zero. By lowering the density
of the central component, we will therefore be able to arrange a slope
that is between zero and that of the surrounding barred bulge.  This
simple reasoning shows that the result of the previous subsection may
be explained if the center of the barred bulge has sufficient central
concentration and is rounder than the outer bar. While if the shape of
the density distribution were to remain independent of radius and only
the central concentration were increased, in first order only the
total number of stars in the central LOS would be changed, which does
not shift the maxima in the histograms.

Fig.~\ref{fig:projections} shows the surface density of the barred
model galaxy in face-on projection, for two slices through the
model. We see from the left panel that the near-planar part of the
boxy bulge-bar shows a significant change in the axis ratio with major
axis length, such that the central parts are nearly round (ellipticity
$\epsilon\sim0.8$) while the outer bar is much more elongated
($\epsilon\sim0.4$).  At the same time, from the spacing of the
contours we see a change of slope in the major axis density profile,
which sets in at approximately major axis length $\simeq 700$ pc
corresponding in projection to $l\simeq 4\dg$. The maxima of the LOS
density distributions and the RC magnitude distributions
overplotted on the surface density clearly show the resulting change
of slope in the inner few degrees.  Note also that the steep 
outer slope on the $l>0$ side is significantly decreased by the volume
effect, and that on the $l<0$ side the outer slope is increased.

The Gaussian dispersions found from the longitude profile fits in
Fig.~\ref{fig:RCsmoohists} at $l=\{-10\dg, -4\dg, -1\dg, 1\dg, 4\dg,
10\dg\}$ are $\{0.28, 0.31, 0.21, 0.20, 0.24, 0.43 \}$ at $b=0\dg$
(Fig.~\ref{fig:RCsmoohists}) and $\{ 0.33, 0.29, 0.21, 0.20, 0.24,
0.40 \}$ at $b=1\dg$, compared to the average VVV values at similar
longitudes and $b=\pm1\dg$ of $\{ 0.41, 0.33, 0.28, 0.30, 0.33, 0.44
\}$.  Averaged over the model longitude profiles for $\vert
l\vert\le4\dg$ and $b=\{ 0\dg, 1\dg\} $, the average Gaussian
$\sigma_{\pm4}\simeq 0.24$ mag, corresponding to a deconvolved FWHM of
0.40 mag or a FWHM in the LOS density distribution of $\pm 750$ pc.
To compare with the VVV data, we need to convolve with the stellar
population broadening $\sigma_{\rm popn}\simeq 0.11$ mag from
Section~\ref{ssec:modelcounts}, and with the broadening by residual
patchy extinction effects within the 3'-6' subfields used for deriving
the reddening map \citep{Gonzalez+11a}. An upper limit for the latter
at $b=\pm1\dg$ can be estimated by comparing the width in J-K colour
of the dereddened RC in a field with A$_{\rm K}=1.5$ with a nearby
low-extinction field at $b\simeq -1$.  The observed difference results
in $\sigma_{\rm J-K}=0.28$ mag (O.~Gonzalez, private communication),
from which $\sigma_{\rm K}\simeq 0.6\sigma_{\rm J-K}=0.17$ mag.
Adding both values in quadrature to the model $\sigma_{\pm4}\simeq
0.24$ mag predicts a total $\sigma=0.31$ mag, which is close to the
observed width for the VVV data, $\sigma_{\pm4}\simeq 0.30$ mag.  The
intrinsic RC widths of the model and of the Galactic bulge as traced
by the VVV RC data may therefore be quite similar.  For comparison,
the projected $\Delta l=\pm4\dg\simeq\pm560 $pc.  Therefore, for both
model and data the structure inside $l=\pm4\dg$ is more extended along
the LOS than in $l$.

By contrast, for the slice with $450 {\rm pc} < \vert z \vert < 750$
pc shown in the right panel of Fig.~\ref{fig:projections} the change
of axis ratio in the barred bulge is much less pronounced, and also
the change of the major axis density slope is less pronounced. As a
result, the density and RC maxima at higher latitudes lie on more
nearly straight lines.

Barred galaxies are often characterized by a rising ellipticity
profile at constant position angle, up to a maximum
\citep[e.g.][]{Peletier+99}, for both disk- and bulge dominated
galaxies \citep{Barazza+08}. N-body models of bars have a range of
morphologies; but isodensities have not been studied extensively. Some
models in \citet{Fux97,Athanassoula+Misiriotis02} appear similar to
the N-body model used above.

\subsection{Total star count map in the inner bulge}
\label{ssec:idea}

\begin{figure}
\includegraphics[angle=-90.0,scale=0.5]{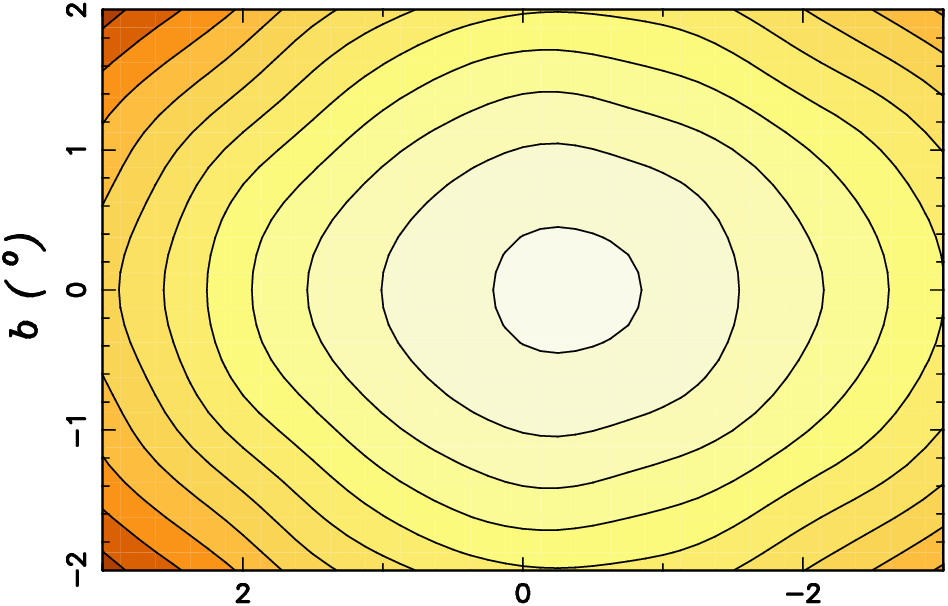}
\includegraphics[angle=-90.0,scale=0.5]{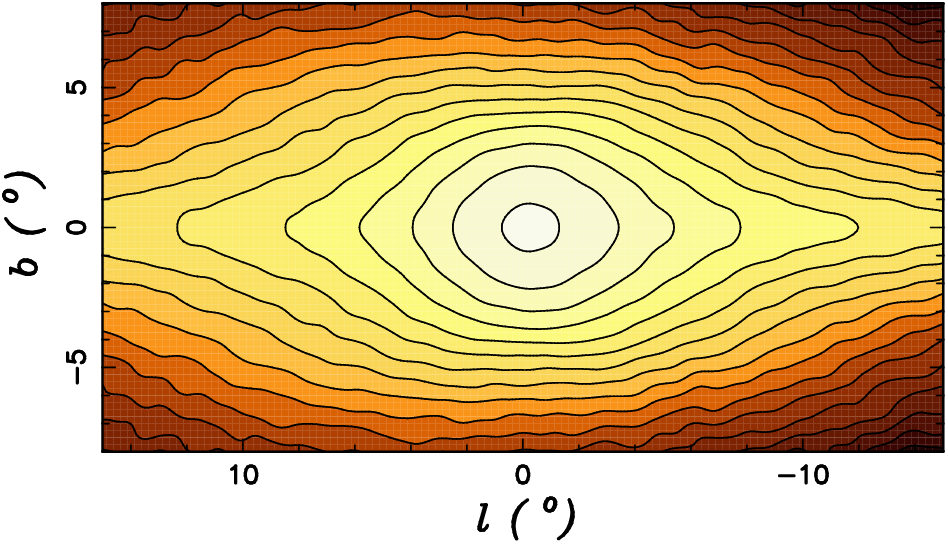}
\caption[]{Map of the particle surface density in a longitude-latitude
  plot as seen by an observer 8 kpc from the model galaxy's
  center. Top: central few degrees; bottom: inner bulge
  region. Contours are logarithmically spaced. The bar angle
  $\alpha_{\rm bar}=25\dg$.}
\label{fig:starcountmap}
\end{figure}

2MASS star count maps for the Galactic bulge in the inner few degrees
show a longitudinal asymmetry \citep[][RFC08]{Alard01} which these
authors interpreted as independent evidence for the possible existence
of a separate nuclear bar.  How does the nuclear bulge of our
simulation compare to these data? Figure~\ref{fig:starcountmap} shows
the smoothed model star count map in $13\dg\times8\dg$ and in the central
$3\dg\times2\dg$. To construct this map, all particles in the
simulation have been used, i.e., possible incompletenesses in the data at the far
side of the bulge are not modelled. We see that the apparent
highest surface density is shifted to $l\simeq-0.3\dg$, as are the
centroids of the surrounding contours. Also the vertical extent of the
isodensities is slightly larger on the $l<0\dg$ side for small $\vert
l\vert$, but becomes larger on the $l>0\dg$ side for larger $\vert
l\vert$.  These asymmetries resemble those predicted by
\citet{Blitz+Spergel91} for the integrated light from the central
region of an inclined triaxial bulge; however, the star counts here
are weighted by two powers of distance more than the integrated light.

The published 2MASS star count maps \citep[][RFC08]{Alard01}
qualitatively show the same asymmetry pattern. The most significant
difference to the model is that the innermost contours in the observed
maps are significantly more flattened than the corresponding model
contours. These innermost 2MASS contours are possibly affected both by
incompleteness (which however would probably decrease the apparent
flattening) and by blending (which might increase it). Similar maps
for the VVV data would therefore be very valuable. Also, the model
used here has a somewhat thicker disk than the Milky Way and, most
importantly, it does not include any dissipative evolution such as gas
infall and star formation on the orbits around the ILR, while such
star formation is clearly on-going in inner Galaxy
\citep{Yusef-Zadeh+09}.  Understanding whether the flattened component
seen in the central 2MASS star counts is consistent with a symmetric
disk-like nuclear bulge \citep{Launhardt+02} or is related to a
nuclear bar will require much more detailed data and analysis. Here we
merely emphasize that the asymmetry structure seen in the 2MASS data
is not a tell-tale evidence for such a secondary bar.

\section{Discussion and conclusions}
\label{sec:discuss}

We have shown that the RC maximum longitude profiles measured for
$b=\pm1\dg$ are well-matched by an N-body model of a boxy bulge and
bar which formed from a bar-unstable disk. This model is the same as
in MVG11, and we have used the same orientation and scaling to avoid
parameter fitting. The change in slope of the model profiles in
  the inner few degrees is caused by a transition from highly
elongated to nearly axisymmetric density contours around this
scale. The average FWHM of the simulated RC profiles in the range
$l=\pm4\dg$ ($\simeq\pm560$ pc at $R_0=8$ kpc) is $\simeq 0.56$ mag,
consistent with the VVV data after convolution with estimated
  stellar population and patchy extinction broadening. This
corresponds to a deconvolved bulge FWHM scale of $\pm 750$ pc along
the LOS in this region. At higher latitudes the transition in the
model isodensities weakens, predicting that the change of slope in the
longitude profile disappears.

The projected star count map of the model has the same asymmetry 
pattern as inferred previously from 2MASS observations. The central
surface density contours in the Galactic bulge are significantly
more flattened, however. This could imply the existence of a
dense central disk around the location of the ILR.

Our results generally support the idea that the main part of the
Galactic bulge formed from the disk through dynamical instabilities. 
They also show that the evidence from star counts so far does not
require a secondary nuclear bar in the Milky Way.

\noindent

\section*{Acknowledgments}

We thank R.~Benjamin, O.~Gonzalez, D.~Minniti, and M.~Rejkuba for
helpful discussions on asymmetries in the GLIMPSE and VVV surveys, the
anonymous referee, and the Aspen Center for Physics for their
hospitality when this work was started.  This paper is based on work
supported in parts by the SPP 1177 program of the German Science
Foundation under Grant GE 567/4-2 and by the US National Science
Foundation under Grant No. 1066293.
 
\bibliographystyle{apj}



\end{document}